\documentclass[aps,prb,twocolumn]{revtex4-1}
\usepackage[utf8]{inputenc}
\usepackage{graphicx,amsmath}
\usepackage{amssymb}
\usepackage{amsthm}
\begin{document}

\title{$\mathbb Z_2$~Green's function topology of Majorana wires}
\author{Jan Carl Budich$^1$ and Bj\"orn Trauzettel$^2$}
\affiliation{$^1$Department of Physics, Stockholm University, Se-106 91 Stockholm, Sweden;\\
 $^2$Institute for Theoretical Physics and Astrophysics,
 University of W$\ddot{u}$rzburg, 97074 W$\ddot{u}$rzburg, Germany}
\date{\today}

\begin{abstract}
 We represent the $\mathbb Z_2$~topological invariant characterizing a one dimensional topological superconductor using a Wess-Zumino-Witten dimensional extension. The invariant is formulated in terms of the single particle Green's function which allows to classify interacting systems. Employing a recently proposed generalized Berry curvature method, the topological invariant is represented independent of the extra dimension requiring only the single particle Green's function at zero frequency of the interacting system. Furthermore, a modified twisted boundary conditions approach is used to rigorously define the topological invariant for disordered interacting systems.
\end{abstract}
\maketitle

\section{Introduction}
The key feature of a topological superconductor (TSC) in one spatial dimension (1D) \cite{Kitaev2001,SauTSC,OppenTSC,BeenakkerReview} is a topologically protected holographic single Majorana bound state (MBS) associated with each of its ends. The recently proposed realization of the 1D TSC in an InSb nanowire with strong spin orbit interaction (SOI) and proximity induced s-wave superconductivity \cite{SauTSC,OppenTSC} is so far the most promising candidate as to its experimental feasibility. Meanwhile, first experimental signatures of MBS have been reported by several experimental groups \cite{LeoMaj, LarssonXu, HeiblumMaj}, however, it should be mentioned that alternative explanations for robust zero bias resonances not owing to Majorana zeromodes have been brought forward \cite{AltlandZero,LeeZero,BeenakkerZero}. Formally, the presence of the MBS is protected by particle hole symmetry (PHS). Note, however, that the PHS in this class of systems does not imply a physical symmetry constraint on the underlying band structure of the normal-conducting spin orbit coupled quantum wire. Rather, the PHS is emergent from the Bogoliubov-deGennes (BdG) mean field description of superconductivity: The BdG band structure consists of two copies of the electronic band structure where the energy spectrum of the hole bands is mirrored as compared to the equivalent electron bands. This enforces the spectrum generating symmetry -- hole bands and electron bands are conjugated by PHS (see Ref. \onlinecite{Budich2013} for a detailed discussion of this point). An additional chiral symmetry present in the ideal model systems proposed in Refs. \onlinecite{SauTSC,OppenTSC} promotes the $\mathbb Z_2$~invariant characterizing the presence of an unpaired MBS to a $\mathbb Z$~invariant \cite{Ryu2002,SauChiral} counting the number of zeromodes at each edge. However, perturbations modifying the SOI as well as magnetic impurities can break the chiral symmetry and gap out paired MBS. The influence of interactions on the topological classification of chiral 1D systems has been analyzed from a matrix product state perspective \cite{Kitaev1DInt}. The classification of a chiral 1D system in terms of its single particle Green's function has also been reported \cite{ZGurarie}. Aditionally, the robustness of the TSC phase in interacting nanowires has been investigated using renormalization group (RG) methods \cite{LossBraunecker,RoschAltlandMaj,FisherInt,LobosLutchyn}.\\

In this work, we extend the original $\mathbb Z_2$~classification in terms of the Pfaffian of the Hamiltonian in Majorana representation reported in Ref. \onlinecite{Kitaev2001} by including both disorder and adiabatic interactions into the classification scheme. We present the $\mathbb Z_2$-classification of the 1D TSC phase which, without additional symmetries, generically belongs to the Cartan-Altland-Zirnbauer class $D$ \cite{AltlandZirnbauer}, in terms of its single particle Green's function. In a first step, we work out explicitly a dimensional extension procedure for a realization of the 1D TSC in a nanowire with strong SOI and proximity induced s-wave superconductivity \cite{SauTSC,OppenTSC}. The dimensional extension allows us to reduce the topological classification of the 1D TSC to that of the 2D $p+ip$~superconductor representing the extended 2D system. The 2D $p+ip$~superconductor is the BdG analog of the quantum anomalous Hall (QAH) effect \cite{QAH} which is characterized by its first Chern number. This procedure fits into the general classification framework of topological field theory (TFT) proposed for time reversal invariant topological insulators in Refs. \onlinecite{QiTFT,
TopologicalOrderParameter}. This framework allows for a reformulation of the invariant in terms of the single particle Green's function. We note that when going beyond mean field the many body Hamiltonian does not show the emergent PHS that is present in the BdG mean field descritption of superconductivity. However, as we discuss in detail below, the single particle Green's function in Nambu space still has a built in PHS \cite{WangTSC} in the presence of arbitrary correlations as an exact feature which is crucial for the construction of the topological invariant via dimensional extension. Upon switching on interactions adiabatically, our classification remains valid for Luttinger liquid like interactions as argued in Ref. \onlinecite{Volovik}. To further simplify the practical calculation of the $\mathbb Z_2$-invariant, we employ a recently proposed generalized Berry curvature method \cite{ZhongWangBerry} which has been applied to 2D and 3D TSC in Ref. \onlinecite{WangTSC}. We show that the interacting invariant can be expressed in terms of the zero frequency single particle Green's function of the physical 1D system, which is independent of the previously introduced extra dimension. Finally, we demonstrate how a hybrid approach of twisted boundary conditions (TBC) \cite{Niu1985} in the physical dimension and periodic boundary conditions in the extra dimension can be used to additionally include disorder at the level of the bulk topological invariant, i.e. without probing the presence of unpaired MBS, quantized zero bias resonances, or other finite size effects. Thus, this approach enables us to topologically classify a 1D TSC in the presence of disorder {\it and} electron-electron interaction.\\    

\section{Model of the 1D TSC}
A lattice model of the 1D TSC \cite{SauTSC,SauChiral} can be cast into the form $H = \int \Psi^\dag \mathcal H_{BdG}\Psi$, where the basis is chosen such that $\Psi=(\psi_{\uparrow},\psi_{\downarrow},\psi_{\uparrow}^\dag,\psi_{\downarrow}^\dag)$. In this basis, the Hamiltonian reads
\begin{eqnarray*}
 \mathcal H_{\rm{BdG}}=\begin{pmatrix}H_0 & \delta \\ \delta^\dag & {-H_0^*}\end{pmatrix}.
\end{eqnarray*}
For the 1D TSC, $H_0(k)= \xi_k + B\sigma_x+u\sin(k)\sigma_y$~is the Hamiltonian of a single channel quantum wire in the presence of a $B$-field induced Zeeman splitting and Rashba SOI. The proximity induced s-wave superconducting gap is of the form $\delta = \begin{pmatrix}0&{-\Delta}\\ \Delta & 0\end{pmatrix}$~and $\xi_k = 1-\cos k-\mu$. Introducing the set of Pauli-matrices $\tau_i$~for the particle hole pseudo spin, the BdG Bloch Hamiltonian reads
\begin{eqnarray}
H_{\rm{BdG}}(k) = \left(\xi_k + B\sigma_x+u \sin(k)\sigma_y\right)\tau_z + \Delta \sigma_y\tau_y.
\label{eqn:ham}
\end{eqnarray}
In this representation, the PHS operation has the intuitive form
\begin{eqnarray}
C = \tau_x K,
\label{eqn:PHS}
\end{eqnarray}
where $K$~denotes complex conjugation. Let us very briefly review the salient physics starting from the continuum model obtained from Eq. (\ref{eqn:ham}) by substituting $\sin(k)\rightarrow k,~\cos(k)\mapsto 1-\frac{k^2}{2}$. For $B=\Delta=0\ne u$, the band structure consists of two particle hole symmetric copies (emergent from the BdG picture) of the shifted Rashba parabolae. The lattice regularization in Eq. (\ref{eqn:ham}) is introduced to make the topological invariants well defined. $\Delta \ne 0$~gaps out the system in its entire Brillouin zone (BZ). For small $k$~this gap competes with a Zeeman gap due to $B\ne 0$~leading to a band inversion at $B^2=\mu^2 + \Delta^2$. For $B^2>\mu^2 + \Delta^2$~we have a TSC with a single MBS associated with each end of a finite wire.\\

\section{Dimensional extension}
Applying the general outline in Ref. \onlinecite{QiTFT}, we explicitly perform a dimensional extension introducing an extra coordinate $v$. Thereby, we connect the 1D TSC state to its 2D parent state, the 2D p+ip superconductor \cite{ReadGreen,Volovik}, which is characterized by its first Chern number. This reduces the topological classification of the noninteracting model to the analysis of the superconducting analog of the QAH effect with chiral Majorana edge states in the extended 2D system. The idea of this procedure is quite simple: Our system cannot be deformed into a trivial 1D superconductor without breaking PHS which provides the topological protection of the TSC phase. However, breaking this constraint, we can deform the TSC, say upon varying $v$~from $0$~to $\pi$, into a trivial 1D superconductor without ever closing the bulk gap of the instantaneous system. It is crucial to perform the particle hole conjugated interpolation to the same trivial state for $v\in \left[-\pi,0\right]$~in such a way that the resulting 2D system is $2\pi$-periodic in the extra coordinate $v$. Then, the extended 2D system is again in the same symmetry class $D$~and its first Chern number $\mathcal C_1$, is well defined independently of our concrete choice of the interpoloation up to even 
integers \cite{QiTFT}. This means that a $\mathbb Z_2$~information, namely $\nu= \mathcal C_1 (\text{mod}~2)$,~is well defined and only depends on the physical 1D system. It is worth noting that finding a suitable interpolation is nontrivial and requires some insight into the physical mechanisms underlying the model. In the following, we will explicitly present an extension which works for a generic 1D TSC and can also be used later on for the disordered interacting system.
The main steps are as follows: First, we switch on a fictitious particle hole breaking gap $\sim \sin(v) \tau_x$~which will keep the gap open for $v\ne 0,\pi~(\text{mod}~2\pi)$. For the 2D system, this term amounts to a $p$-wave superconducting pairing whereas the intermediate 1D systems are only formally defined because they break the emergent PHS symmetry of the BdG picture. However, the procedure is still well defined since the physical information is only encoded in the original 1D system. During the deformation, the band inversion is destroyed by a term $\sim\beta(2-2\cos v)\sigma_y\tau_y$~which vanishes for the physical model ($v=0$). For sufficiently large $\beta$~this term will produce a trivial superconducting phase for $v=\pm \pi$~as it enhances the superconducting gap by $4\beta$~beyond the critical strength $\Delta_c=\sqrt{B^2-\mu^2}$.
In summary, the Wess-Zumino-Witten (WZW) \cite{WZW} extended Hamiltonian reads
\begin{eqnarray}
\mathcal H_{\rm{WZW}}(k,v)= H_{\rm{BdG}}(k)+\sin(v) \tau_x+\beta(2-2\cos v)\sigma_y\tau_y.
\end{eqnarray}
Integrating the Berry curvature $\mathcal F$~of this Hamiltonian over the $(k,v)$~BZ indeed yields
\begin{eqnarray}
\mathcal C = \frac{1}{2\pi}\int_{BZ} \mathcal F = \theta(B^2-\Delta^2-\mu^2),
\end{eqnarray}
valid for parameters close enough to the band inversion that no artificial level crossings which depend on the details of the lattice regularization occur.\\
\begin{figure}
\begin{minipage}{0.99\linewidth}
\includegraphics[width=\linewidth]{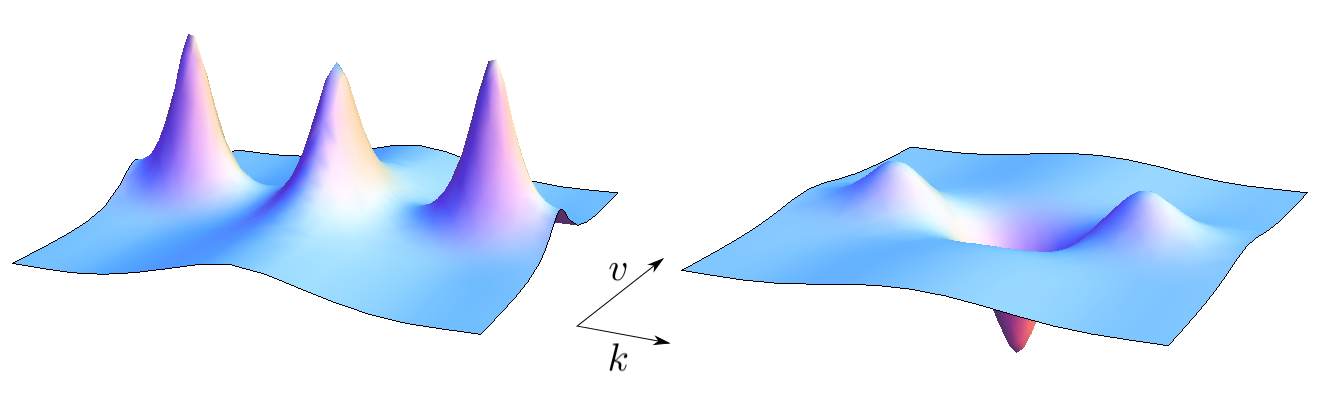}
\end{minipage}
\caption{(Color online) 3D plot of the Berry curvature $\mathcal F(k,v)$~for nontrivial parameters ($B=1.5$) left and trivial parameters ($B=0.4$) right. $u=\Delta=2\beta=1,~\mu=0$~in both plots.}
\label{fig:cleancomp}
\end{figure}
In Fig. \ref{fig:cleancomp}, we compare the Berry curvature of an extension of a nontrivial 1D TSC  with that of a trivial superconducting wire. In the extra dimension $v$, the modulus of the curvature is smoothly decaying without any notable difference between the trivial and the nontrivial case. This is reflected in our derivation below, which shows that the topological invariant of the translation-invariant system can be defined in terms of its single particle Green's function without reference to the extra dimension. Note that this picture changes in the framework of TBC as introduced to account for the presence of disorder as is discussed below and shown in Fig. \ref{fig:discomp}.\\

\section{Single particle Green's function topology}
We now discuss the possiblity to include interactions into the proposed classification scheme by generalizing the Chern number of the noninteracting system to a topological invariant of the single particle Green's function in combined frequency momentum space, as has been proposed for time reversal invariant topological insulators in 2D and 3D \cite{TopologicalOrderParameter}. In even spatial dimension $2n$, it has been shown \cite{Redlich1984, VolovikQH3HE,Golterman1993} that the homotopy of the single particle Green's function determines the Hall conductance $\sigma_{xy}$~($n=1$) and its higher dimensional analogues, respectively, of an insulating fermionic system. It has been explicitly demonstrated \cite{QiTFT} that this representation of $\sigma_{xy}$~adiabatically connects to the $n$th-Chern number associated with the Berry curvature of a Bloch Hamiltonian in the non-interacting limit. The adiabatic assumption in this case implies that the gapped ground state of the interacting system can be 
continuously connected to the manifold of occupied single particle levels of the noninteracting system which is also separated from the empty states by a finite energy gap. The most prominent example of interactions which are non-adiabatic in the sense just defined lead to the $\frac{1}{\nu}$~fractional quantum Hall (FQH) effect in a partially filled lowest Landau level (LLL). In this case, the noninteracting system is gapless and the single particle states that span the LLL split into $\nu$~degenerate ground states when TBC are applied \cite{Niu1985}. This ground state degeneracy is at the basis of the concept of topological order \cite{WenTO} which rigorously classifies FQH systems.

In 1D, interactions play a peculiar role generically leading to non-Fermi-liquid behavior. From a viewpoint of perturbation theory interactions are therefore considered to be non-adiabatic in 1D, as no meaningful quasiparticles can be defined. However, it has been argued \cite{Volovik} that the Fermi surface properties as described by the momentum space topology of the single particle Green's function are still adiabatically connected to those of the non-interacting system.\\

Rewritten in terms of the single particle Green's function $G(i\omega,p)$~with $p=(k,v)$~of the extended system, i.e., $G_0(i\omega,p)= \left(i \omega-\mathcal H_{\rm{WZW}}(p)\right)^{-1}$~for the special case of the non-interacting system, the $\mathbb Z_2$~invariant $\nu$~reads
\begin{eqnarray}
\nu = \frac{\epsilon^{\mu\nu\rho}}{24 \pi^2}\int_{BZ\times \mathbb R_\omega}\text{Tr}\left[GG^{-1}_\mu GG^{-1}_\nu GG^{-1}_\rho\right]~(\text{mod}~2),
\label{eqn:volovik}
\end{eqnarray}
where $G^{-1}_\mu= \partial_\mu G^{-1},~\mu = 0,1,2=\omega,k,v$, and $\mathbb R_\omega$~denotes the frequency axis. Recently, it has been generally shown \cite{ZhongWangBerry} that for an interacting system, an invariant of this form can be simplified by introducing a generalized Berry curvature $\mathcal {\tilde F}= -i \sum_{R-zeros}(d\langle p,\alpha\rvert)\wedge d\lvert p,\alpha\rangle$~associated with the fictitious non-interacting Hamiltonian $\tilde H= -G^{-1}(0,p)$, which takes into account the eigenvectors $\lvert p,\alpha\rangle$~of $G^{-1}(0,p)$~with positive eigenvalues, the so called R-zeros \cite{ZhongWangBerry}. Here, $\alpha$~labels the internal degrees of freedom of the single particle Green's function like orbital labels and spin. The $\mathbb Z_2$~invariant then takes the form of a generalized Chern number, i.e.
\begin{eqnarray}
\nu = \frac{1}{2\pi}\int_{BZ} \mathcal {\tilde F}~(\text{mod}~2).
\end{eqnarray}
As has been demonstrated for the noninteracting case in Ref. \cite{QiTFT} the $\mathbb Z_2$~classification of the particle hole symmetric 1D system can then be further simplified to
\begin{eqnarray}
\nu = 2 P(0) (\text{mod}~2)=\frac{1}{\pi}\int_{0}^{2\pi}dk \mathcal {\tilde A}(k)~(\text{mod}~2),
\label{eqn:pol}
\end{eqnarray}
where $P(0)$~is the polarization of the physical 1D system and $\mathcal{\tilde A} (k)=-i \sum_{R-zeros}\langle k,\alpha\rvert \partial_k \lvert k,\alpha\rangle$~is the generalized Berry connection restricted to the physical system at $v=0$, i.e. at $p=(k,0)$. Note that this general form does no longer depend on the dimensional extension procedure and can be calculated once the zero frequency single particle Green's function $G(0,k)$~is known. Finally, the $Z_2$-invariant can be practically determined by formal analogy to the non-interacting case by calculating the Majorana number \cite{Kitaev2001} defined in terms of the Pfaffian of the fictitious non-interacting Hamiltonian $\tilde H$~in Majorana representation.\\

We would like to point out some subtleties in our analysis stemming from the fact that we are dealing with a superconducting system here rather than a normal insulator. Most importantly, it is not obvious how the emergent PHS present in the BdG mean field description translates to a many body Hamiltonian which has terms that are quartic in the field operators. More concretely, a standard Hubbard interaction term $U \Psi_\uparrow^\dag \Psi_\uparrow \Psi_\downarrow^\dag \Psi_\downarrow$~does not change sign under the particle hole conjugation $\psi\rightarrow \psi^\dagger$. However, the generalization of our dimensional extension procedure to interacting systems only relies on a built in particle hole symmetry for the single particle Green's function and not for the many body Hamiltonian. In the absence of interactions, the PHS operation (\ref{eqn:PHS}) of the mean field Hamiltonian (\ref{eqn:ham}) can be expressed for the free Nambu Green's function as
\begin{align}
\tau_x G(i\omega,k)\tau_x=-G^T(-i\omega,-k).
\end{align}
In Ref. \onlinecite{WangTSC}, it has been shown using the spectral representation that this PHS emergent from the Nambu spinor representation of the Green's function remains an exact feature for the Nambu Green's function associated with an arbitrary many-body Hamiltonian. This observation renders our dimensional extension arguments well defined in the presence of interactions.\\

Furthermore, we note that the physical interpretation of the invariant $\nu$~as a polarization in Eq. (\ref{eqn:pol}) does not imply immediate observable consequences since it represents a polarization of Bogoliubov quasiparticles that do not have a well defined electric charge. Along similar lines, the Chern number of the 2D $p+ip$~superconductor is known not to be in one to one correspondence to a quantized Hall conductivity.

\section{Disorder and Twisted Boundary Conditions}
Our formulation so far has been relying on translation-invariance which implies the existence of a BZ. This description will thus no longer be applicable in the presence of disorder. To this end, the concept of TBC has been introduced to topologically classify quantum Hall systems in the absence of translation-invariance \cite{Niu1985}. As long as a bulk mobility gap is present, the Green's function is exponentially bounded in real space for energies in this gap. Under these conditions, Niu {\it et al.} \cite{Niu1985} showed that the Hall conductance can be represented as a constant ground state Berry curvature with the wave vector replaced by the twisting angles $\theta,\phi$ of the TBC. In this formalism, the Hall conductance $\sigma_{xy}$~reads
\begin{eqnarray}
&\sigma_{xy}=2\pi i G_0\left(\langle \frac{\partial\psi_0}{\partial\theta }\rvert\frac{\partial\psi_0}{\partial\phi}\rangle -\langle \frac{\partial\psi_0}{\partial\phi}\rvert\frac{\partial\psi_0}{\partial\theta}\rangle\right)\equiv \nonumber\\
&2\pi i G_0 \mathcal F_{\theta \phi}=
G_0 \int_{T^2}\frac{i \mathcal F_{\theta \phi}}{2\pi},
\label{eqn:Niu}
\end{eqnarray}
where $\psi_0$~denotes the ground state wave function, $G_0=\frac{e^2}{h}=\frac{1}{2\pi}$~is the quantum of conductance. In the last equality of Eq. (\ref{eqn:Niu}), the independence of $\mathcal F_{\theta\phi}$~on the twisting angles which is the main result of Ref. \onlinecite{Niu1985} has been used to make the topological quantization of $\sigma_{xy}$~manifest by representing it as $G_0$~times the Chern number of the $U(1)$-bundle over the torus $T^2$~of the twisting angles $(\theta,\phi)$.\\

Since in this work, we consider a disordered 1D system, we can without loss of generality assume translation-invariance in the extra dimension. Integrating over the momentum $v$~associated with the direction of translational invariance is equivalent to evaluating Eq. (\ref{eqn:Niu}) for $\theta=0$~for the special case of a system with translational invariance in $x$-direction. We therefore introduce a hybrid approach of twisted boundary conditions in the physical dimension and $v$-momentum integration in the extra dimension. The mixed Chern number of the extended 2D system in the presence of this stripe-like disorder can be expressed as
\begin{eqnarray}
\mathcal C = \int_{-\pi}^{\pi} dv\int_{-\pi}^{\pi}d\phi~\frac{\mathcal {\hat F}_{v\phi}}{2\pi} = \int_{-\pi}^{\pi} dv ~\mathcal {\hat F}_{v\phi}
\label{eqn:mixed}
\end{eqnarray}
where $\mathcal {\hat F}$~is the Berry curvature on the mixed torus defined by the wave vector in $v$-direction and the twisting angle $\phi$~of the TBC imposed in the physical direction. The first equality sign in Eq. (\ref{eqn:mixed}) makes the integer quantization of our topological invariant manifest by formal analogy to the familiar Chern number in momentum space. The second equality sign in Eq. (\ref{eqn:mixed}) follows from the independence of $\int_{-\pi}^{\pi}dv\mathcal {\hat F}_{v\phi}= \left.\mathcal F_{\theta\phi}\right |_{\theta=0}$~of the twisting angle $\phi$~which is a direct consequence of the independence of $\mathcal F_{\theta\phi}$~on both $\theta$~and $\phi$.\\

The main advantage of Eq. (\ref{eqn:mixed}) as compared to the general 2D case (see Eq. (\ref{eqn:Niu})) is that only the eigenstates of a 1D system have to be calculated to evaluate the topological invariant which is numerically less costly. This program allows for a topological classification of disordered systems with periodic boundaries, i.e., without explicitly probing the presence of unpaired MBS or performing a disorder average to effectively reestablish translation-invariance. By employing the generalized Berry curvature $\mathcal {\tilde F}$~defined below Eq. (\ref{eqn:volovik}) in terms of the single particle Green's function, this procedure allows for a direct calculation of the topological invariant of a system in the presence of both interactions and disorder.\\

For non-interacting systems with closed boundary conditions, the influence of disorder on the 1D TSC phase has been carefully studied before on the basis of a scattering matrix approach \cite{SmatrixDisorderMaj,BrouwerPRBDisorder,BrouwerDisorder}. Thereby, the situation of single disorder configurations as well as disorder averaging has been considered. Certain relations between the localization length and the superconducting coherence length have been identified that led to a detailed characterization of the phase diagram for Majorana wires in the presence of disorder. Interestingly, disorder can even induce a topological phase under specific conditions in Majorana wires \cite{InancDisorder}. We demonstrate below that our new approach also yields a meaningful result for the topological invariant in the presence of disorder. Our intention is not to verify the previously discovered phase diagram of the disordered Majorana wire with our method. Rather, we would like to illustrate how our method distinguishes at the level of the bulk topological invariant defined in Eq. (\ref{eqn:mixed}) two examples where disorder does and does not lead to a topological phase transition.\\

\begin{figure}
\centering
\begin{minipage}{0.7\linewidth}
\includegraphics[width=\linewidth]{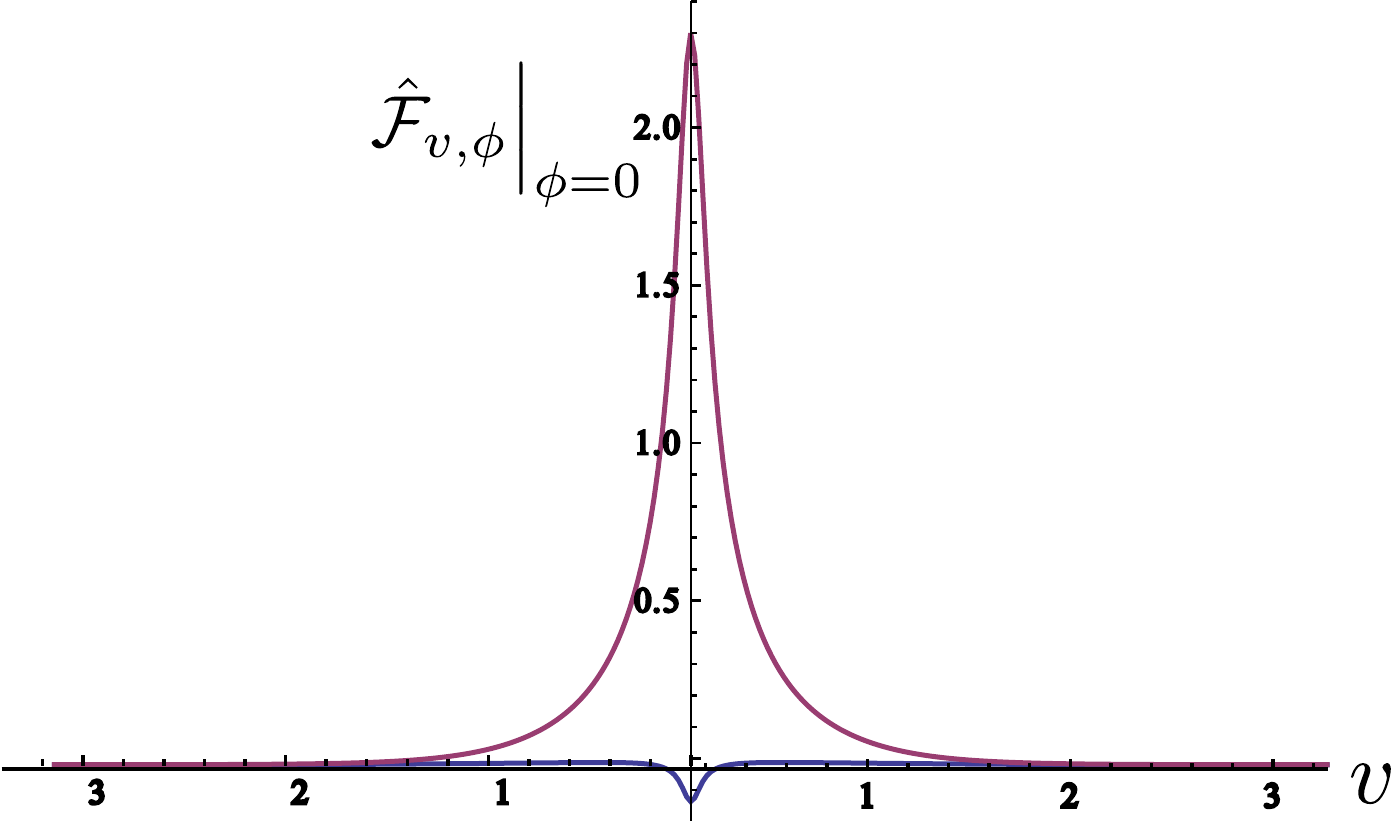}
\end{minipage}
\caption{(Color online) Topological invariant for two Majorana wires (see Eq. (\ref{eqn:ham})) with different disorder strength $\gamma$. Plot of the Berry curvature $\left.\mathcal{\hat F}_{v,\phi}\right|_{\phi=0},$~for nontrivial parameters ($\Delta=0.7,~\gamma=u=B=\beta=1,~\mu=0$) (purple) and trivial parameters ($~\Delta=0.7,~\gamma=5,~u=B=\beta=1,~\mu=0$) (blue). The wire length is chosen to be 100 sites in both plots. The value of $\mathcal C$ is given by the integral over the curves, i.e., $\mathcal C=1$~for the purple and $\mathcal C=0$~for the blue curve, respectively.}
\label{fig:discomp}
\end{figure}

Let us give an example for the calculation of our mixed Chern number. In Fig. \ref{fig:discomp}, we show the mixed Berry curvature $\left.\mathcal{\hat F}_{v,\phi}\right|_{\phi=0}$~for a weakly disordered system ($\gamma =1$) in the topologically nontrivial phase and a strongly disordered trivial system ($\gamma=5$). Here, $\gamma$~is the strength of a scalar Gaussian onsite potential $V(x)$, i.e., $\langle V(x)V(x')\rangle_{\text{disorder}}=\gamma^2\delta(x-x')$. Note that in contrast to the translation-invariant case (see Eq. (\ref{eqn:pol})), the topology is determined by the $v$-dependence of the mixed Berry curvature. The mixed Chern numbers of the two exemplary systems shown in Fig. \ref{fig:discomp} are just given by the integration of the plotted mixed Berry curvature $\mathcal{\hat F}$~over $v$, as defined in Eq. (\ref{eqn:mixed}). The two plots clearly show systems which due to different disorder strengths have different mixed Chern numbers. We would like to point out that our approach is well defined for a single realization of disorder and does not involve a disorder-average.\\

Finally, we would like to point out that even for $\gamma =1$, the onsite potential fluctuations significantly exceed the bulk insulating gap of the 1D TSC in our example. The disorder-induced transition from nontrivial to trivial takes place at disorder strengths which are, depending on the other model parameters typically three to five times larger than the bulk gap which is in agreement with recent results obtained from level spectroscopy in 1D TSC with closed boundary conditions \cite{FranzDisorder}. Note however, that we have only considered short ranged 
disorder as measured by the Fermi wavelength. The wave-function can self-average on the length scale of the Fermi wavelength which considerably weakens the effect of disorder on the level spectrum. Hence, we expect the robustness of the topological phase to be also dependent on the correlation length of disorder and not only on its strength.    
\\

\section{Conclusions}
We have constructed a dimensional extension which reduces the topological classification of the 1D TSC phase to calculating the first Chern number of the 2D parent state, the p+ip superconductor. This approach is ready-made to rephrase the invariant in terms of the single particle Green's function of the extended system. Using a generalized Berry curvature method, the invariant can be simplified to the noninteracting classification scheme with a fictitious non-interacting Hamiltonian defined in terms of the Green's function of the physical 1D system at zero frequency. To obtain a well defined topological invariant for disordered systems in the presence of interactions, we have introduced a hybrid approach of twisted boundary conditions in the physical dimension and momentum integration in the extra dimension.\\   

\section*{Acknowledgments}
We would like to thank Dietrich Rothe, Ronny Thomale, Michael Wimmer, and Shoucheng Zhang for interesting discussions and acknowledge financial support from the DFG-JST Research Unit "Topotronics" (BT and JCB) as well as from the Swedish Science Research Council (JCB).\\

%
\end{document}